\documentclass[11pt]{article}
\usepackage{amsmath, amssymb, amsthm}
\usepackage{geometry}
\usepackage{mathpazo}
\usepackage{microtype}
\usepackage{titlesec}
\usepackage{hyperref}
\usepackage{xcolor}
\usepackage{ulem}   
\newcommand{\ind}{1\hspace{-2.1mm}{1}}
\newcommand{\eps}{\varepsilon}
\newcommand{\half}{\frac{1}{2}}

\newcommand{\D}{\mathrm{d}}
\newcommand{\EE}{\mathbb{E}}

\newtheorem{assumption}{Assumption}
\newtheorem{remark}{Remark}
\newtheorem{theorem}{Theorem}

\definecolor{darkblue}{RGB}{0,0,128}

\hypersetup{
  colorlinks=true,
  linkcolor=darkblue,
  citecolor=darkblue,
  urlcolor=darkblue,
  pdftitle={
    Local Stochastic Rough Volatility:
    Pathwise Filtering and the Conditional Density Equation
  },
  pdfauthor={Damiano Brigo and Vladimir Lucic},
  pdfsubject={
    Conditional-density SPDEs, stochastic flows,
    rough volatility and Rao--Blackwellized calibration
  },
  pdfkeywords={
    rough Heston, local stochastic volatility,
    Ito-Wentzell, stochastic flow, Fokker-Planck,
    nonlinear filtering, Rao-Blackwellization
  }
}

\title{\textbf{Local Stochastic Rough Volatility: Pathwise Filtering
and the Conditional Density Equation}}
\author{Damiano Brigo \ \  \  \  Vladimir Lucic \\
Department of Mathematics\\ Imperial College London}
\date{22 Aug 2026}

\begin{document}

\maketitle

\begin{abstract}
This article studies the conditional-density equation and its pathwise
transformation in local stochastic rough volatility models, with rough
Heston (rHeston) as the main explicit example. Under the stated
common-filtration, measurability, predictability and spatial-regularity
assumptions, we show that the It\^o--Wentzell random-PDE reduction of the
conditional density SPDE remains valid under local stochastic rough
volatility. After fixing a common-environment realization and the associated
stochastic flow, the transformed equation becomes a deterministic PDE with path-dependent coefficients. This
yields a pathwise Fokker--Planck formulation that connects naturally with
Rao--Blackwellized calibration. In the pure rough Heston case, the
transformed coefficients simplify and the conditional density admits an
explicit lognormal form.
\end{abstract}

\medskip


\medskip



\section{Introduction}

The main result of this article is that, under the stated common-filtration,
predictability, coefficient-regularity and conditional-density regularity
assumptions, the common-noise transport term in the conditional-density
SPDE for local stochastic rough-volatility models can be removed by an
It\^o--Wentzell stochastic-flow transformation. Before the common
environment is fixed, the transformed equation is a random
Fokker--Planck-type equation; for each fixed common-environment
realisation, it is an ordinary deterministic divergence-form PDE with
coefficients that may be merely measurable in time. Thus, no rough
differential term remains explicitly after the transformation. Rough
volatility enters through the path dependence and temporal irregularity of
the stochastic flow and of the transformed coefficients, rather than
through an additional rough transport operator. This result has been inspired by 
the pathwise nonlinear filtering results pioneered by Martin Clark.

A closely related approach is developed by Bank, Bayer, Friz and
Pelizzari~\cite{bank2025rough}. They condition general, possibly
non-Markovian LSV dynamics on the Brownian motion driving volatility,
formulate the conditional state dynamics as a rough stochastic differential
equation, and characterize conditional European prices through a backward
linear RPDE and a rough Feynman--Kac representation. Our emphasis is
different but complementary: we study the forward conditional law. Starting
from the conditional-density SPDE, we use the It\^o--Wentzell flow to
conjugate away the common-noise transport and obtain a forward pathwise
Fokker--Planck equation. The distinction is therefore between a backward
rough-path-stable pricing representation and a forward flow-transformed
density representation, rather than between incompatible descriptions of
the conditional dynamics. In their additive- and multiplicative-SV examples,
Bank et al.\ also reduce the corresponding RPDEs to ordinary heat-type
equations and recover conditional Gaussian or lognormal laws. In the present
article, the same lognormal structure appears as the fundamental solution of
the forward flow-transformed density equation and enters directly into the
Rao--Blackwellized leverage calibration ratio. The forward-density
formulation also connects directly to Rao--Blackwellized nonlinear filtering
and the associated leverage-function calibration update, providing a direct
algorithmic framework for fitting LSRV models to implied-volatility surfaces.
In pure rough Heston, it further makes the conditional lognormal density
explicit.

The conditional-density viewpoint for local-stochastic-volatility (LSV)
models leads naturally to a stochastic partial differential equation in
which the common noise appears through a first-order transport term. In the
classical diffusion setting, this transport term can be removed by a random
change of variables generated by the associated stochastic flow. The present
article studies how this picture extends when the volatility factor is rough or,
more generally, Volterra-driven and non-Markovian.

Recent related work also uses It\^o--Wentzell techniques in volatility
modelling from a different angle, with emphasis on skew; see
Fukasawa~\cite{fukasawa2026skew} for a recent preprint in that direction.

Recent developments in stochastic filtering and rough-path formulations are
also relevant for context. Crisan and Pardoux~\cite{CrisanPardoux2026}
study filtering equations in spaces of measures for general correlated
signal--observation systems, allowing observation-dependent coefficients and
degenerate observation diffusion, and establish uniqueness links between the
Zakai and Kushner--Stratonovich equations. From a rough-path viewpoint,
Bugini, Friz, L\^e and Zhang~\cite{BuginiFrizLeZhang2026} develop rough
counterparts of the Zakai and Kushner--Stratonovich equations and prove their
well-posedness, together with a randomization procedure connecting the rough
and stochastic formulations. Related rough Fokker--Planck developments for
common-noise McKean--Vlasov systems are studied by Bugini, Friz and
Stannat~\cite{BuginiFrizStannat2025}. In a different but adjacent SPDE
direction, Das, Guo and Loeper~\cite{DasGuoLoeper2025} derive conditional
Feynman--Kac representations through backward SPDEs arising in derivative
pricing. More broadly, rough-path methods have also been used directly at the
level of rough-volatility model construction; see, for example, Bonesini,
Ferrucci, Gasteratos and Jacquier~\cite{BonesiniFerrucciGasteratosJacquier2026}.
These works are complementary to the present article. Our focus here is not the
general well-posedness theory of filtering equations in spaces of measures,
nor the direct rough-path construction of rough-volatility dynamics, but the
conditional-density SPDE arising in local stochastic rough volatility and its
pathwise It\^o--Wentzell reduction to a deterministic Fokker--Planck type PDE
with path-dependent coefficients.

The main structural point is that the roughness of the volatility factor does
not itself obstruct the It\^o--Wentzell transformation, provided the common
driver $Z$ remains a Brownian motion in the chosen common filtration and the
relevant coefficients and conditional density satisfy the stated
measurability and spatial-regularity assumptions. In that case the
common-noise transport can still be absorbed into a stochastic flow, and
after fixing the common environment the conditional-density equation becomes
a deterministic Fokker--Planck type PDE with path-dependent
coefficients.\footnote{More general continuous-semimartingale drivers could
be treated after modifying the SPDE according to their finite-variation and
quadratic-variation characteristics.}

Lucic~\cite{Lucic1} addresses the classical LSV diffusion setting, deriving
the conditional forward equation, stochastic Dupire-type equations and
related It\^o--Wentzell identities in that non-rough framework. The present
article uses that classical work as background but makes a different
contribution: it studies local stochastic rough volatility, formulates the
conditional-density SPDE under a common rough/Volterra environment, proves
pathwise cancellation of the common-noise transport term, and derives the
corresponding deterministic pathwise Fokker--Planck equation, with rough
Heston as an explicit example. Accordingly, the present article should be
read as a rough-volatility extension of the classical LSV picture rather
than as a reformulation of it. 

The present paper gives a self-contained formulation for local
stochastic rough volatility, states the standing assumptions directly at the
level needed here, derives the transformed pathwise PDE, and specializes the
construction to rough Heston.

The paper is organized as follows. Section~\ref{sec:LSRV} introduces the
general LSRV framework, the common environment, and the conditional-density
SPDE. Section~\ref{sec:flow_reduction} derives the stochastic-flow
transformation and the It\^o--Wentzell cancellation, leading to the
transformed pathwise PDE. Section~\ref{sec:interpretation_calibration}
discusses interpretation, well-posedness, regularity, numerical implications
and the Rao--Blackwellized calibration ratio. Section~\ref{sec:rH_easier}
specializes the framework to rough Heston, where the flow simplifies and the
conditional density can be written explicitly.

\section{General LSRV setting and conditional-density SPDE}
\label{sec:LSRV}

\subsection{Rough volatility and rough Heston}\label{sec:rHestonintro}

We have abundant evidence that traditional SVMs cannot reproduce the
volatility surfaces observed in real markets. As a response, SVMs involving a
fractional Brownian motion with Hurst exponent in $(0,\half)$ have been
proposed. In Gatheral, Jaisson and Rosenbaum~\cite{gatheral2018volatility},
such a model was introduced based on statistical observations under the
historical measure, and triggered a large research response that has extended
the boundaries of such models and their applications. The overall consensus
coming from these and related papers is that, while rough volatility behaviour
can be mimicked by standard Markovian models, \emph{`rough volatility as the
null statistical hypothesis is very hard to beat'} (quoting the paper of
Gatheral, Jaisson and Rosenbaum). We briefly recall the rough Heston model,
which will later serve as our main explicit specialization.

The rHeston variance is driven by a Brownian motion through a Volterra
kernel, producing a non-Gaussian process with fractional Brownian-like local
regularity.

A fractional Brownian motion~$W^H$ is a continuous-time centered Gaussian
process starting from zero, with covariance function
\[
\EE\!\left[W_{t}^{H} W_{s}^{H}\right]
= \tfrac{1}{2}\!\left(|t|^{2H}+|s|^{2H}-|t-s|^{2H}\right),
\qquad\text{for all }s,t\geq 0.
\]
Here, $H\in(0,\tfrac{1}{2})$ is called the Hurst parameter and describes the
roughness of the paths of the process, in the sense of H\"older
$(H-\eps)$ regularity for any $\eps\in(0,H)$, and $H=\half$ corresponds to
the standard Brownian motion. Such a process can be built on top of a
two-sided standard Brownian motion~$W$ via the Mandelbrot--Van Ness
representation
\begin{equation}\label{eq:MandelbrotfBm}
W_{t}^{H}
= \frac{c_H}{\Gamma(H+\half)}
\left(
\int_{-\infty}^{0}
\!\left[(t-s)^{H-\half}-(-s)^{H-\half}\right]\D W_{s}
+\int_{0}^{t}(t-s)^{H-\half}\D W_{s}
\right),
\end{equation}
almost surely for all $t\geq 0$, where~$\Gamma$ denotes the Gamma function.
Here $c_H$ is a normalising constant chosen so that
$\EE[(W_t^H)^2]=t^{2H}$; see Decreusefond and
\"{U}st\"{u}nel~\cite{decreusefond1999} for details of this normalisation
and the associated filtration properties. Here we just recall that
in this normalization one may take
\[
c_H = \bigl[\Gamma(2H+1)\sin(\pi H)\bigr]^{1/2},
\]
so that the full prefactor in~\eqref{eq:MandelbrotfBm} is
\[
\frac{c_H}{\Gamma(H+\half)}
=
\frac{\bigl[\Gamma(2H+1)\sin(\pi H)\bigr]^{1/2}}
{\Gamma(H+\half)}
=
\left(
\frac{2H\,\Gamma(\tfrac32-H)}
{\Gamma(H+\tfrac12)\Gamma(2-2H)}
\right)^{1/2}.
\]
Without this normalization, the construction gives a process with variance
$\tilde c_H\,t^{2H}$ for a constant $\tilde c_H\neq 1$ in general.

Other fractional Gaussian processes and alternative Volterra representations
are used in applications. On a finite time interval, standard fractional
Brownian motion admits a Volterra representation of the form
\begin{equation}\label{eq:fBmVolterra}
W_t^H=\int_0^t K_H(t,s)\,\D W_s,
\end{equation}
for a suitable deterministic Volterra kernel $K_H(t,s)$. By contrast, the
Riemann--Liouville (or type~II) fractional process is obtained from the
convolution kernel $(t-s)^{H-\half}\ind_{\{s<t\}}$, namely
\begin{equation}\label{eq:RLfBm}
\widetilde W_t^H=\int_0^t (t-s)^{H-\half}\,\D W_s.
\end{equation}
The local singularity of the kernel near the diagonal governs the H\"older
regularity. For the Riemann--Liouville kernel, the associated Volterra
process is adapted to the Brownian filtration, and under the standard
nondegeneracy properties of this kernel one has filtration equivalence with
the driving Brownian motion; see Decreusefond and
\"Ust\"unel~\cite{decreusefond1999}. For a general square-integrable Volterra
kernel, such filtration equivalence requires additional invertibility
assumptions and need not hold in general.

Inspired by this convolutional representation, the rough Heston model reads
\begin{align*}
\frac{\D S_{t}}{S_{t}}
&= \sqrt{V_{t}}\,\D B_{t},
\qquad S_0>0,\\
V_{t}
&= V_{0}+\frac{1}{\Gamma\!\left(H+\tfrac{1}{2}\right)}
\int_{0}^{t}(t-s)^{H-\frac{1}{2}}
\!\left\{\kappa(\theta-V_{s})\,\D s
+\nu\sqrt{V_{s}}\,\D W_{s}\right\},
\end{align*}
under the risk-neutral measure (where we ignore interest rates and dividends
for simplicity), where~$S$ denotes the stock price process, $V$ the
instantaneous variance process, and~$B$ and~$W$ are two standard Brownian
motions with correlation $\rho\in[-1,1]$, the ``leverage parameter'', and
with $V_0,\kappa,\theta,\nu>0$ and $H\in(0,\half)$. We will assume the
``factor'' structure
\begin{equation}\label{eq:factor}
\D W_t = \D Z_t,\qquad
\D B_t = \rho\,\D Z_t+\sqrt{1-\rho^2}\,\D\zeta_t,
\end{equation}
for the two Brownian motions, where $Z$ and $\zeta$ are two independent
standard Brownian motions. Since $\D W_t=\D Z_t$, the volatility driver is
contained in the common environment; in the rough Heston setting we take the
common filtration to be generated by $(Z,V)$, with $Z$ serving as the common
Brownian driver in the conditional-density SPDE. The cross-variation
$\D\langle B,W\rangle_t=\rho\,\D t$ follows immediately from the factor
structure~\eqref{eq:factor}.

This is the rough Heston model. One important benefit of adopting this
rHeston model is that the affine property of the standard Heston model is
preserved (in an infinite-dimensional setting), so that pricing European
options can be achieved in principle via Fourier transform methods; see
Lewis~\cite{lewis2000} for the general inversion framework and El~Euch and
Rosenbaum~\cite{eleuch2019characteristic} for the rough Heston
characteristic function via the fractional Riccati equation. However, the
evaluation of the characteristic function via the fractional Riccati equation
can be numerically delicate in some market regimes. In particular, for short
maturities and deep out-of-the-money strikes, predictor--corrector schemes
may exhibit stability issues, and the oscillatory integrals arising in
Fourier inversion can become challenging in practice. In such regimes one may
prefer Monte Carlo methods. More generally, many path-dependent claims are
not directly accessible through the standard terminal characteristic function
and are therefore commonly treated by Monte Carlo, although selected affine
path functionals may still admit transform-based formulas.

Rough Heston is a stochastic rough-volatility model without a local
component. We now pass to the more general local stochastic rough volatility
setting, of which rough Heston is a particular example.

\begin{remark}[Common filtration and conditioning]\label{rem:filtration}
We work on a stochastic basis satisfying the usual conditions. Let
\[
\widehat{\mathcal C}_t
:= \sigma\bigl(Z_s, V_s : 0\le s\le t\bigr)\vee\mathcal N,
\]
completed and made right-continuous, where $\mathcal N$ denotes the
$\mathbb P$-null sets. We assume that $Z$ is a
$\widehat{\mathcal C}_t$-Brownian motion and that $\zeta$ is independent of
$\widehat{\mathcal C}_\infty$. 

For every $t>0$, we assume that the conditional law of $S_t$ given
$\widehat{\mathcal C}_t$ is absolutely continuous on $[0,\infty)$ with
density $p_t$, in the sense that for every Borel set $A\subseteq[0,\infty)$,
\[
  \mathbb P(S_t\in A\mid\widehat{\mathcal C}_t)
  =
  \int_A p_t(s)\,\D s.
\]
At $t=0$, the conditional law is
\[
  p_0=\delta_{S_0}
\]
in the sense of measures.

All pathwise statements below are understood for $\mathbb{P}$-almost
every common-environment realization $(Z_{[0,T]},V_{[0,T]})$. In
particular, conditioning is always on $\widehat{\mathcal C}_t$ (information
up to time $t$), and the pathwise analysis for a fixed realisation of $(Z,V)$
implicitly uses the full path; this is the standard convention in the
Clark--Crisan framework and is made explicit here for precision.
Finally, although a complete environment path is fixed when analysing the PDE
on $[0,T]$, its coefficient at time $t$ depends only on the common
information available up to time $t$.
\end{remark}

One final important point is that for this model, we work with a continuous
nonnegative weak solution $V$ of the Volterra square-root equation on the
chosen stochastic basis, satisfying the common-filtration compatibility
assumption of Remark~\ref{rem:filtration}.

\subsection{General local stochastic rough volatility}

Rough volatility models with a local component, which we call local
stochastic rough volatility (LSRV) models, take the form
\begin{equation}\label{eq:LSRV}
\D S_t = (r_t-d_t)\,S_t\,\D t + S_t\,L(t,S_t)\,f(V_t)\,\D B_t,
\end{equation}
where $V_t$ is generated by a rough-volatility or Volterra-volatility model,
such as rough Heston or rough Bergomi, the noise being built on a Brownian
motion $W_t$. The factor structure~\eqref{eq:factor} is kept throughout.

For the conditional-density SPDE derived below, we introduce the
following notation for the key coefficient functions. The total instantaneous
volatility is $\Sigma_t(s) := s\,L(t,s)\,f(V_t)$, so that the total
instantaneous variance is $\Sigma_t^2(s) = s^2 L^2(t,s) f^2(V_t)$. The
common-noise loading is $\Gamma_t(s) := \rho\,\Sigma_t(s) =
\rho\,s\,L(t,s)\,f(V_t)$, and the idiosyncratic variance is
$\Delta_t^2(s) := (1-\rho^2)\,s^2\,L^2(t,s)\,f^2(V_t)$, so that
$\Sigma_t^2(s) = \Gamma_t^2(s) + \Delta_t^2(s)$. These definitions are used
throughout Sections~\ref{sec:LSRV}--\ref{sec:rH_easier}.

The approach of reducing a stochastic equation to a deterministic one via a
robust representation was pioneered for the multiplicative-noise Zakai
equation by Clark~\cite{clark1978} and further formalized by Clark and
Crisan~\cite{clark_crisan2005}. The extension of this philosophy to handle
the first-order transport noise observed in the Wentzell equation relies
heavily on the stochastic characteristic flows developed by
Kunita~\cite{kunita1990}. In the present article we use this framework to
analyze the conditional-density SPDE in LSRV and to connect the transformed
equation with local-leverage calibration.

\begin{assumption}[Coefficient measurability and spatial regularity]
\label{ass:coeffs}
On the stochastic basis of Remark~\ref{rem:filtration}, assume that the
random fields
\[
(t,\omega,s)\mapsto \Gamma_t(\omega,s),\qquad
\Delta_t^2(\omega,s),\qquad
\mu_t(\omega,s)
\]
are $\widehat{\mathcal P}(\widehat{\mathcal C})\otimes\mathcal
B([0,\infty))$-measurable, where $\widehat{\mathcal
P}(\widehat{\mathcal C})$ denotes the predictable $\sigma$-field of the
filtration $(\widehat{\mathcal C}_t)$. The spatial regularity and growth
conditions below hold for $\mathbb P$-almost every $\omega$.
\begin{enumerate}
\item[\textup{(i)}] The map $s\mapsto\Gamma_t(s)
      :=\rho\,s\,L(t,s)\,f(V_t(\omega))$ belongs to $C^3(\mathbb{R}_+)$
      uniformly in $t\in[0,T]$, with $\partial_s^k\Gamma_t$ bounded for
      $k=1,2,3$, and with $\Gamma_t$ of at most linear growth in~$s$.
\item[\textup{(ii)}] The idiosyncratic variance
      $\Delta_t^2(\cdot)$ belongs to $C^2_{\mathrm{loc}}([0,\infty))$
      and the drift $\mu_t(\cdot)$ belongs to
      $C^1_{\mathrm{loc}}([0,\infty))$, both uniformly in $t\in[0,T]$.
      For every $R<\infty$,
      \[
        \sup_{t\in[0,T]}\sup_{0\leq s\leq R}
        \left|\partial_s^k\Delta_t^2(s)\right|<\infty,
        \quad k=0,1,2,
      \]
      and
      \[
        \sup_{t\in[0,T]}\sup_{0\leq s\leq R}
        \left|\partial_s^k\mu_t(s)\right|<\infty,
        \quad k=0,1.
      \]
      Moreover, $\Delta_t^2$ has at most quadratic growth and $\mu_t$
      has at most linear growth in $s$, uniformly in $t$.
\end{enumerate}
\end{assumption}

\begin{assumption}[Common-environment path regularity]
\label{ass:environment}
The path $t\mapsto V_t(\omega)$ is locally bounded and (Borel) measurable
on $[0,T]$ for $\mathbb P$-almost every $\omega$.
\end{assumption}

\begin{assumption}[Conditional-density semimartingale field]
\label{ass:density}
For every $\tau > 0$, $p$ is an adapted semimartingale random field on
$[\tau,T]\times(0,\infty)$ admitting the decomposition
\begin{equation}\label{eq:p_semimartingale}
  p_t(s) = p_\tau(s)
         + \int_\tau^t a_u(s)\,\D u
         + \int_\tau^t g_u(s)\,\D Z_u,
  \qquad t\in[\tau,T],
\end{equation}
where $a$ and $g$ are progressively measurable with respect to
$(\widehat{\mathcal C}_t)$ and, for every compact
$K\Subset(0,\infty)$,
\[
  p_t(\cdot)\in C^2(K),\qquad
  a_t(\cdot)\in C^0(K),\qquad
  g_t(\cdot)\in C^1(K),
\]
with
\[
  \sup_{t\in[\tau,T]}\|p_t\|_{C^2(K)}<\infty
  \quad\text{a.s.},
\]
and
\[
  \int_\tau^T
  \left(
    \|a_u\|_{C^0(K)}
    +\|g_u\|_{C^1(K)}^2
  \right)\,\D u<\infty
  \quad\text{a.s.}
\]
The initial condition $p_0 = \delta_{S_0}$ is understood in the sense of
measures. In the general LSRV result, the existence of a
conditional-density random field satisfying~\eqref{eq:p_semimartingale}
on every $[\tau,T]$, $\tau>0$, is a standing assumption.
\end{assumption}

\begin{assumption}[Nondegenerate density regime]
\label{ass:rho}
Throughout the density-based analysis of
Sections~\ref{sec:LSRV}--\ref{sec:conclusion} we impose
\begin{equation}\label{eq:rho_strict}
  |\rho| < 1.
\end{equation}
The boundary case $|\rho|=1$ is discussed at the end of
Remark~\ref{rem:wellposed} below and requires a measure-valued
reformulation.
\end{assumption}

\begin{remark}[Interpretation of the standing assumptions]
\label{rem:assumptions_interpretation}
Assumptions~\ref{ass:coeffs}--\ref{ass:density} are the standing hypotheses
used throughout the density-based analysis.
Assumption~\ref{ass:coeffs} is the coefficient-level regularity required for
the stochastic flow and for the transformed coefficients to be locally
spatially smooth. Any global growth estimates additionally require the
corresponding global bounds on the stochastic flow and its spatial
derivatives. Assumption~\ref{ass:environment} records the temporal
regularity actually needed from the common-environment path. Assumption~\ref{ass:density}
encodes the semimartingale random-field structure required to apply the
It\^o--Wentzell formula. In the present SPDE setting one has
\[
  g_t(s) = -\partial_s\bigl(\Gamma_t(s)p_t(s)\bigr),
  \qquad
  a_t(s) = \tfrac{1}{2}\partial_{ss}\bigl(\Sigma_t^2(s)p_t(s)\bigr)
         - \partial_s\bigl(\mu_t(s)p_t(s)\bigr),
\]
where these identifications are consistent with the
SPDE~\eqref{eq:SPDE_main} and are understood whenever the stated
local classical derivatives exist. They are consequences of the
SPDE structure under the regularity imposed above, not additional
hypotheses.
\end{remark}

\begin{remark}[Verification in rough Heston]
\label{rem:rough_heston_assumptions}
For rough Heston specifically, Assumption~\ref{ass:coeffs}(i) holds since
$L\equiv1$ and $\Gamma_t(s)=\rho s\sqrt{V_t}$ is affine in~$s$, hence
belongs to $C^3(\mathbb{R}_+)$ with bounded derivatives of orders $1,2,3$
and linear growth. Assumption~\ref{ass:environment} holds almost surely
since $V$ has continuous sample paths and is therefore locally bounded.
Assumption~\ref{ass:coeffs}(ii) is satisfied in the rough Heston case:
$\Delta_t^2(s) = (1-\rho^2)\,s^2 V_t$ belongs to
$C^\infty_{\mathrm{loc}}(\mathbb{R}_+)$ with derivatives bounded on every
compact spatial subset, and has at most quadratic growth in $s$;
$\mu_t(s)=(r_t-d_t)\,s$ is affine in $s$, hence
$C^\infty_{\mathrm{loc}}$ with linear growth, assuming $V$ is continuous
on $[0,T]$ and $r-d$ is bounded. In the general LSRV setting,
Assumption~\ref{ass:coeffs}(ii) must be verified separately for the chosen
local-volatility function $L(t,s)$.
Assumption~\ref{ass:density} replaces the classical $C^{2,3}$ regularity
requirement with the appropriate semimartingale random-field formulation:
since $t\mapsto p_t(s)$ is driven by the It\^o integral against $Z$, it
has Brownian temporal regularity and cannot be classically $C^1$ in time.
The spatial regularity $p_t(\cdot)\in C^2_{\mathrm{loc}}$ and
$g_t(\cdot)\in C^1_{\mathrm{loc}}$, together with the local continuity
and integrability conditions on $a$ and $g$ stated above, are intended to
provide a standard sufficient framework for applying the It\^o--Wentzell
formula (Kunita~\cite{kunita1990}, Theorem~3.3.1). We assume throughout
that $r$, $d$ and $L$ are deterministic Borel functions.
\end{remark}

In the present local stochastic rough-volatility setting, the spot price is
driven by a Brownian motion that decomposes into a common-noise component
and an orthogonal idiosyncratic component, while the volatility factor
$V_t$ is part of the common environment. In this notation the common noise
is denoted by $Z_t$ and enters the conditional-density equation directly.

For completeness, we briefly record the weak derivation of the
conditional-density SPDE. Let $\varphi\in C_c^2((0,\infty))$ be a smooth
test function, and write
\[
\langle p_t,\varphi\rangle := \int_0^\infty \varphi(s)\,p_t(s)\,\D s.
\]
Conditional on the common environment, the spot dynamics have drift
$\mu_t(s)$, common-noise coefficient $\Gamma_t(s)$, and total
instantaneous variance
\[
\Sigma_t^2(s)=\Gamma_t^2(s)+\Delta_t^2(s).
\]
Hence the conditional law $p_t$ satisfies
\[
\D\langle p_t,\varphi\rangle
=
\left\langle p_t,\,
\mu_t\,\varphi'
+\tfrac{1}{2}\Sigma_t^2\,\varphi''
\right\rangle \D t
+
\left\langle p_t,\Gamma_t\,\varphi'\right\rangle \D Z_t.
\]

After localization, the relevant stochastic integrals are square
integrable. Independence of $\zeta$ from
$\widehat{\mathcal C}_\infty$, together with the Brownian property of
$\zeta$ in the full filtration, gives
\[
  \mathbb E\!\left[
    \int_0^t
    \sqrt{1-\rho^2}\,S_uL(u,S_u)f(V_u)\,\varphi'(S_u)\,\D\zeta_u
    \,\middle|\,
    \widehat{\mathcal C}_t
  \right]
  =0.
\]
The optional-projection identity for stochastic integrals with respect
to the $\widehat{\mathcal C}_t$-Brownian motion $Z$ gives
\[
  {}^o\!\left(
    \int_0^t
    \Gamma_u(S_u)\varphi'(S_u)\,\D Z_u
  \right)
  =
  \int_0^t
  \mathbb E\!\left[
    \Gamma_u(S_u)\varphi'(S_u)
    \mid\widehat{\mathcal C}_u
  \right]\D Z_u.
\]

Since
\[
  \mathbb E\!\left[
    \Gamma_u(S_u)\varphi'(S_u)
    \mid \widehat{\mathcal C}_u
  \right]
  =
  \langle p_u,\Gamma_u\,\varphi'\rangle,
\]
this yields precisely the $Z$-integral displayed above.

Integrating by parts in the spatial variable then yields the
divergence-form SPDE in the distributional sense; under
Assumption~\ref{ass:density} it holds locally in the stated classical
spatial sense:
\begin{align}\label{eq:SPDE_main}
\D p_{t}(s)
&= -\partial_{s}\!\bigl(\Gamma_{t}(s)p_{t}(s)\bigr)\,\D Z_{t} \nonumber\\
&\quad
+\left\{
\tfrac{1}{2}\partial_{ss}\!\bigl(\Sigma_{t}^{2}(s)p_{t}(s)\bigr)
-\partial_{s}\!\bigl(\mu_{t}(s)p_{t}(s)\bigr)
\right\}\D t,
\end{align}
where $\mu_t(s)=(r_t-d_t)s$, and the coefficients $\Sigma_t(s)$,
$\Gamma_t(s)$, $\Delta_t^2(s)$ are as defined above,
with $\rho\in(-1,1)$ throughout the density-based analysis. The SPDE is
driven by $\D Z_t$ (not by the full spot driver $\D B_t$) because we
condition on the common environment $(Z,V)$; the spot driver decomposes as
$\D B_t=\rho\,\D Z_t+\sqrt{1-\rho^2}\,\D\zeta_t$, and the total
instantaneous variance splits as
\begin{equation}\label{eq:var_split}
\Sigma_t^2(s) = \underbrace{\Gamma_t^2(s)}_{\text{common-noise part}}
              + \underbrace{\Delta_t^2(s)}_{\text{idiosyncratic part}},
\qquad
\Delta_t^2(s) := (1-\rho^2)\,s^2 L^2(t,s)f^2(V_t).
\end{equation}
The $Z$-driven transport is removed by the stochastic flow constructed
below; only the $\zeta$-driven idiosyncratic diffusion $\Delta_t^2$ remains
in the transformed PDE.

\section{Stochastic flow and It\^o--Wentzell reduction}
\label{sec:flow_reduction}

The central idea of the pathwise filtering framework is to absorb the
stochastic transport term
$-\partial_{s}(\Gamma_{t}(s)p_{t}(s))\,\D Z_{t}$ into a random change of
coordinates.

Let $\Phi_t$ be the stochastic flow generated by the common-noise transport
coefficient. For the $i$-th common-environment realization, the flow
satisfies:
\begin{equation}\label{eq:flow}
\D\Phi_{t}^{i}(x)
= \rho\,\Phi_{t}^{i}(x)\,L(t,\Phi_{t}^{i}(x))\,f(V_{t}^{i})\,\D Z_{t}^{i},
\qquad \Phi_{0}^{i}(x)=x.
\end{equation}

\begin{remark}[Flow properties]\label{rem:flow}
Several structural properties of the flow~\eqref{eq:flow} are recorded here
for precision.
\begin{enumerate}
\item[\textup{(i)}] \emph{Predictability.}
      The coefficient field $\Gamma_t(s)$ is predictable with respect to
      the common filtration $(\widehat{\mathcal C}_t)$ by
      Assumption~\ref{ass:coeffs}. In the structural representation
      $\Gamma_t(s)=\rho\,s\,L(t,s)\,f(V_t)$, this is consistent with the
      common-filtration setup of Remark~\ref{rem:filtration}; in the rough
      Heston case, $V$ is $\widehat{\mathcal C}_t$-adapted with continuous
      paths, hence predictable by the standard result that adapted
      left-continuous (or continuous) processes are predictable. This is
      the measurability condition required by Kunita~\cite{kunita1990} for
      the stochastic-flow construction to apply.
\item[\textup{(ii)}] \emph{Invariance of $\mathbb{R}_+$.}
      Since $\Gamma_t(0) = 0$ for all $t\in[0,T]$, the point $x=0$ is a
      fixed point of the flow~\eqref{eq:flow}, and $\mathbb{R}_+$ is
      invariant under $\Phi_t$. This allows Kunita's flow theorem, which
      is stated on $\mathbb{R}^d$, to be applied after extending $\Gamma_t$
      smoothly and consistently to $\mathbb{R}$; the extension preserves
      the $C^3$ spatial regularity and linear-growth bound of
      Assumption~\ref{ass:coeffs}(i), and the flow restricted to $\mathbb{R}_+$
      coincides with the solution of~\eqref{eq:flow}.
\item[\textup{(iii)}] \emph{Positivity of the Jacobian.}
      Differentiating the flow equation~\eqref{eq:flow} with respect to
      the initial condition $x$ gives the variational
      equation~\eqref{eq:Jt_ito} for $J_t(x)=\partial_x\Phi_t(x)$. This
      is a linear SDE with the explicit solution
      \begin{equation}\label{eq:Jt_explicit}
        J_t(x)
        = \exp\!\left(
            \int_0^t \Gamma_u'(\Phi_u(x))\,\D Z_u
            - \tfrac{1}{2}\int_0^t
              \bigl(\Gamma_u'(\Phi_u(x))\bigr)^2\,\D u
          \right) > 0
        \quad\text{a.s.}
      \end{equation}
      The strict positivity of $J_t(x)$ confirms that $\Phi_t$ is
      orientation-preserving, so that the change of variables
      $q_t(x) = p_t(\Phi_t(x))\,J_t(x)$ uses $J_t$ rather than $|J_t|$,
      and the mass-conservation identity holds without absolute values.
\item[\textup{(iv)}] \emph{Transformed initial condition.}
      Since $\Phi_0(x)=x$ and $J_0(x)=1$, the transformed density
      satisfies $q_0 = (\Phi_0^{-1})_{\#}p_0 = p_0 = \delta_{S_0}$.
      With deterministic initial stock price $S_0$, the initial
      conditional density is $p_0 = \delta_{S_0}$ in the sense of
      measures, and accordingly $q_0 = \delta_{S_0}$.
      Assumption~\ref{ass:density} is imposed on every interval
      $[\tau,T]$, $\tau>0$.
\item[\textup{(v)}] \emph{Interpretation of $\partial_t q$.}
      When the coefficients of the transformed pathwise PDE are only
      measurable in $t$ (as is the case under Assumption~\ref{ass:environment}),
      the time derivative $\partial_t q$ is understood almost everywhere
      in $t$, or in the sense of distributions, rather than as a classical
      pointwise derivative.
\item[\textup{(vi)}] \emph{Order of construction.}
      The stochastic flow and the It\^o--Wentzell identity are constructed
      on the original probability space before any pathwise freezing. Only
      after these stochastic identities are established on a common
      full-probability set do we freeze the common environment and
      interpret the resulting coefficients as deterministic path-dependent
      coefficients. In particular, once a single environment path is
      frozen, the resulting $Z$-path is just a deterministic sample path
      rather than a Brownian motion in its own right. 
\end{enumerate}
\end{remark}

The validity of transforming the SPDE into a pathwise PDE relies
fundamentally on the It\^o--Wentzell formula (Kunita~\cite{kunita1990},
Theorem~3.3.1), which requires the spatial regularity of $p_t$ stated in
Assumption~\ref{ass:density}.

Before freezing the common environment, $Z$ is a standard Brownian
motion in the common filtration, so the It\^o calculus required to construct
the stochastic flow $\Phi$ applies in the usual way. After the stochastic
identities have been established, one may then freeze a common-environment
path and regard the resulting coefficients as deterministic functions of
time. The fractional nature of $V_t^i$ does not invalidate
equation~\eqref{eq:flow}; rather, once the $i$-th joint path scenario is
fixed, $V_t^i$ simply acts as a deterministic, albeit highly irregular and
non-differentiable, time-dependent input to the flow (see
Assumption~\ref{ass:environment}). In general, roughness in time of the volatility
coefficient is not an obstruction; what matters is the spatial regularity of
$\Gamma_t$ as a function of $s$, which is controlled by
Assumption~\ref{ass:coeffs}(i).

Assuming $\Phi_t$ is a diffeomorphism (guaranteed by
Assumption~\ref{ass:coeffs}(i) via Kunita~\cite{kunita1990}), we define the
Jacobian $J_t(x)=\partial_x\Phi_t(x)$ and the second derivative
$H_t(x)=\partial_{xx}\Phi_t(x)$. Differentiating the It\^o flow with respect
to the initial spatial coordinate $x$ gives the variational equations
\begin{align}
\D\Phi_t(x) &= \Gamma_t(\Phi_t)\,\D Z_t, \label{eq:flow_short}\\
\D J_t(x)   &= \Gamma_t'(\Phi_t)\,J_t\,\D Z_t, \label{eq:Jt_ito}\\
\D H_t(x)   &= \bigl[\Gamma_t''(\Phi_t)\,J_t^2
                + \Gamma_t'(\Phi_t)\,H_t\bigr]\D Z_t.
\label{eq:Ht_ito}
\end{align}
No bounded-variation term appears in $\D J_t$: the differentiation is with
respect to the spatial parameter $x$, not an application of It\^o's formula
in time to a nonlinear function of $\Phi_t$.

\subsection{Main result: the transformed pathwise PDE}
\label{sec:main_result}

For the weak-initial-trace statement below, define
\[
  C_{S_0}([0,T];(0,\infty))
  :=
  \bigl\{x\in C([0,T];(0,\infty)):x_0=S_0\bigr\}.
\]

\begin{theorem}[Flow-transformed conditional-density equation]
\label{thm:main}
Let Assumptions~\ref{ass:coeffs}--\ref{ass:density}  hold and let $\Phi_t$ be the stochastic flow
generated by
\[
  \D\Phi_t(x) = \Gamma_t(\Phi_t(x))\,\D Z_t, \qquad \Phi_0(x)=x.
\]
For every $\tau>0$, define
\[
  q_t(x) := p_t(\Phi_t(x))\,J_t(x), \qquad t\in[\tau,T].
\]
Define further
\[
  \widetilde\Delta_t^2(x)
  =
  \frac{\Delta_t^2(\Phi_t(x))}{J_t(x)^2},
\]
and
\[
  \tilde\mu_t(x)
  =
  \frac{\mu_t(\Phi_t(x))}{J_t(x)}
  -\frac12
   \frac{\Delta_t^2(\Phi_t(x))H_t(x)}{J_t(x)^3}.
\]
Then, for $\mathbb P$-almost every common-environment realisation,
$q_t$ satisfies
\[
  \partial_t q_t
  =
  -\partial_x(\tilde\mu_tq_t)
  +\tfrac12\partial_{xx}(\widetilde\Delta_t^2 q_t)
\]
on $[\tau,T]\times(0,\infty)$, with coefficients
\eqref{eq:Delta_tilde}--\eqref{eq:mu_tilde}.
The identity is understood pointwise almost everywhere in time under
the stated local classical regularity, and otherwise in the
distributional sense.
Under the additional nondegeneracy condition
Assumption~\ref{ass:rho}, the residual diffusion coefficient
$\widetilde\Delta_t^2$ is nontrivial and the transformed equation
remains in the density-based regime considered in the sequel.

If, in addition, there exist a standard Borel path space $E$ and a
Borel measurable, nonanticipative map
\[  \mathcal S:
  E\times C_0([0,T];\mathbb R)
  \longrightarrow
  C_{S_0}([0,T];(0,\infty))
\]
such that, with
\[
  \mathcal E:=(Z_{[0,T]},V_{[0,T]}),
\]
one has
\[
  S=\mathcal S(\mathcal E,\zeta)
  \qquad\text{indistinguishably},
\]
and $\zeta$ is independent of $\mathcal E$, then
$(q_t)_{t>0}$ has the weak initial trace
\[
  q_t\Longrightarrow\delta_{S_0}
  \qquad\text{as }t\downarrow0.
\]
\end{theorem}

We now justify the weak initial trace in the second part of
Theorem~\ref{thm:main}. 
Let $E$ be the standard Borel path space from the theorem, let
$\mathbb W$ denote Wiener measure on $C_0([0,T];\mathbb R)$, and write
\[
  \mathcal E := (Z_{[0,T]},V_{[0,T]}).
\]
For each admissible environment path $e\in E$, define the transformed
path-valued map
\[
  X^e(\eta)
  :=
  \bigl(\Phi^e\bigr)^{-1}\!\bigl(\mathcal S(e,\eta)\bigr),
  \qquad
  \eta\in C_0([0,T];\mathbb R),
\]
where $\mathcal S:E\times C_0([0,T];\mathbb R)\to
C_{S_0}([0,T];(0,\infty))$ is the nonanticipative solution map from the
theorem and $\Phi^e$ denotes the flow
associated with the frozen environment $e$. Let
\[
  Q^e := \mathbb W\circ (X^e)^{-1}
\]
be the induced probability measure on $C([0,T];(0,\infty))$, and let
$Q_t^e$ denote its time-$t$ marginal. By construction, $Q_t^e$ is the law of
$X_t^e$ when the idiosyncratic Brownian path is sampled under Wiener measure
and the common environment is frozen at $e$.

Because $\mathcal S$ is nonanticipative and $\zeta$ is independent of
$\widehat{\mathcal C}_\infty$, the time-$t$ marginal $Q_t^{\mathcal E}$
is a version of the conditional law of
\[
  X_t := \Phi_t^{-1}(S_t)
\]
given $\widehat{\mathcal C}_t$. Since $q_t$ is precisely the conditional
density of $X_t$ given $\widehat{\mathcal C}_t$, it follows that for
$\mathbb P$-almost every environment realisation $e$ and every bounded
continuous test function $\varphi\in C_b((0,\infty))$,
\[
  \int_0^\infty \varphi(x)\,q_t^e(x)\,\D x
  =
  \int_{C([0,T];(0,\infty))}\varphi(\xi_t)\,Q^e(\D \xi)
  =
  \int_{C_0([0,T];\mathbb R)}\varphi\bigl(X_t^e(\eta)\bigr)\,\mathbb W(\D \eta).
\]
Now $X^e$ has continuous paths and
\[
  X_0^e = S_0,
\]
because $\Phi_0^e=\mathrm{id}$ and $\mathcal S(e,\eta)\in
C_{S_0}([0,T];(0,\infty))$. Therefore,
for every fixed admissible $e$ and every $\eta$,
\[
  X_t^e(\eta)\to S_0
  \qquad\text{as }t\downarrow0.
\]
By dominated convergence,
\[
  \int_0^\infty \varphi(x)\,q_t^e(x)\,\D x
  \longrightarrow
  \varphi(S_0)
  \qquad\text{as }t\downarrow0,
\]
for $\mathbb P$-almost every $e$. Hence
\[
  q_t^e \Longrightarrow \delta_{S_0}
  \qquad\text{as }t\downarrow0,
\]
which is the asserted weak initial trace.

In a Monte Carlo implementation, for the $i$-th fixed common-environment
realization we write
\begin{equation}
q_{t}^{i}(x) := p_{t}^{i}(\Phi_{t}^{i}(x))\,J_{t}^{i}(x),
\end{equation}
and the transformed density $q_t^i$ solves the pathwise deterministic
equation
\begin{equation}\label{eq:pathwise_pde}
\partial_{t}q_{t}^{i}(x)
= -\partial_{x}\!\bigl(\tilde{\mu}_{t}^{i}(x)\,q_{t}^{i}(x)\bigr)
+ \tfrac{1}{2}\partial_{xx}\!\left(
  \widetilde{\Delta}_{t}^{2,i}(x)\,q_{t}^{i}(x)
\right),
\end{equation}
where the transformed variance and drift coefficients are:
\begin{align}
\widetilde{\Delta}_{t}^{2,i}(x)
&= (1-\rho^{2})\,
   \frac{(\Phi_{t}^{i}(x))^{2}\,L^{2}(t,\Phi_{t}^{i}(x))\,f^{2}(V_{t}^{i})}
        {(J_{t}^{i}(x))^{2}},
\label{eq:Delta_tilde}\\
\tilde{\mu}_{t}^{i}(x)
&= (r_t-d_t)\,\frac{\Phi_{t}^{i}(x)}{J_{t}^{i}(x)}
 - \tfrac{1}{2}(1-\rho^{2})\,
   \frac{(\Phi_{t}^{i}(x))^{2}\,L^{2}(t,\Phi_{t}^{i}(x))\,
         f^{2}(V_{t}^{i})\,H_{t}^{i}(x)}
        {(J_{t}^{i}(x))^{3}}.
\label{eq:mu_tilde}
\end{align}
Equation~\eqref{eq:pathwise_pde} is, before fixing the common environment,
a random Fokker--Planck type PDE with path-dependent coefficients; after
fixing the common environment and the associated stochastic flow, it becomes
a deterministic PDE without an explicit rough driver. Under additional
nondegeneracy assumptions it is parabolic in the classical sense. The
``roughness'' introduced by the rHeston model has been structurally
localized: it no longer appears as a stochastic transport term in the SPDE,
but instead manifests through rough, path-dependent coefficients in the
transformed equation. The proof of Theorem~\ref{thm:main} follows.

\subsection{Proof of Theorem~\ref{thm:main}: exact cancellation and
transformed coefficients}\label{sec:Proof1}

We seek to show the exact cancellation of the transport noise and derive the
explicit forms of the transformed drift $\tilde{\mu}_t$ and variance
$\widetilde{\Delta}_t^2$ under
Assumptions~\ref{ass:coeffs}--\ref{ass:density}. Fix $\tau>0$. We apply the
It\^o--Wentzell formula on $[\tau,T]$ to $p_t(\Phi_t(x))$. Since $\tau>0$ is
arbitrary, the resulting equation holds on $(0,T]$. Under the additional
hypotheses in the second part of Theorem~\ref{thm:main}, the family also
satisfies
\[
  q_t\Longrightarrow\delta_{S_0}
  \qquad\text{as }t\downarrow0.
\]
The original Wentzell SPDE is:
\begin{equation}\label{eq:spde_start}
\D p_t(s)
= -\partial_s(\Gamma_t(s)p_t)\,\D Z_t
+ \left[\tfrac{1}{2}\partial_{ss}(\Sigma_t^2(s)p_t)
        -\partial_s(\mu_t(s)p_t)\right]\D t.
\end{equation}
We split the total variance as in~\eqref{eq:var_split}:
\begin{align*}
  \Sigma_t^2(s) &= \Gamma_t^2(s)+\Delta_t^2(s),\\
  \Delta_t^2(s) &= (1-\rho^2)s^2L^2(t,s)f^2(V_t),
\end{align*}
where $\Delta_t^2$ is the idiosyncratic variance.

Let us evaluate the density along the flow:
\[
Y_t(x):=p_t(\Phi_t(x)).
\]
By the It\^o--Wentzell formula (Kunita~\cite{kunita1990}, Thm.~3.3.1),
applicable under Assumption~\ref{ass:density}, the differential is:
\begin{equation}\label{eq:ito_w_full}
\D Y_t
= (\D p_t)\big|_{\Phi_t}
+ (\partial_s p_t)\big|_{\Phi_t}\D\Phi_t
+ \tfrac{1}{2}(\partial_{ss}p_t)\big|_{\Phi_t}\D\langle\Phi\rangle_t
+ \langle\D(\partial_s p_t),\D\Phi\rangle_t.
\end{equation}

\medskip
\noindent\textbf{Step 1: The stochastic terms.}
Isolating the $\D Z_t$ terms from~\eqref{eq:ito_w_full} and applying the
product rule to the transport operator
$-\partial_s(\Gamma_t p_t)=-\Gamma_t'p_t-\Gamma_t\partial_s p_t$:
\begin{align}
(\D Y_t)_{\mathrm{stoch}}
&= \bigl[-\Gamma_t'(\Phi_t)p_t(\Phi_t)
         -\Gamma_t(\Phi_t)\partial_s p_t(\Phi_t)\bigr]\D Z_t
  +\partial_s p_t(\Phi_t)\,\Gamma_t(\Phi_t)\,\D Z_t \nonumber\\
&= -\Gamma_t'(\Phi_t)\,p_t(\Phi_t)\,\D Z_t
 = -\Gamma_t'(\Phi_t)\,Y_t(x)\,\D Z_t.
\end{align}

\medskip
\noindent\textbf{Step 2: The drift terms.}
The cross-variation term requires computing the stochastic part of
$\D(\partial_s p_t)$. Differentiating the transport term in
\eqref{eq:spde_start} with respect to $s$ gives the martingale part
$\D(\partial_s p_t)\big|_{\mathrm{mart}} = -\partial_{ss}(\Gamma_t
p_t)\,\D Z_t$. Taking the bracket with
$\D\Phi_t=\Gamma_t(\Phi_t)\,\D Z_t$:
\begin{equation}
\langle\D(\partial_s p_t),\D\Phi\rangle_t
= -\partial_{ss}(\Gamma_t p_t)\big|_{\Phi_t}\cdot\Gamma_t(\Phi_t)\,\D t
= -\Gamma_t(\Phi_t)\,(\Gamma_t p_t)_{ss}(\Phi_t)\,\D t.
\end{equation}
Collecting all $\D t$ terms from~\eqref{eq:ito_w_full}:
\begin{equation}
(\D Y_t)_{\mathrm{drift}}
= \left[\tfrac{1}{2}(\Sigma_t^2 p_t)_{ss}
        -(\mu_t p_t)_s
        +\tfrac{1}{2}p_{ss}\Gamma_t^2
        -\Gamma_t(\Gamma_t p_t)_{ss}\right]\D t.
\end{equation}
Substituting $\Sigma_t^2=\Gamma_t^2+\Delta_t^2$ and grouping all $\Gamma_t$
terms:
\begin{equation}\label{eq:gamma_group}
\tfrac{1}{2}(\Gamma_t^2 p_t)_{ss}
+\tfrac{1}{2}\Gamma_t^2 p_{ss}
-\Gamma_t(\Gamma_t p_t)_{ss}.
\end{equation}
Expanding the spatial derivatives using the product rule:
\begin{align*}
(\Gamma_t p_t)_{ss}
&= \Gamma_t''p_t+2\Gamma_t'\partial_s p_t+\Gamma_t p_{ss},\\
(\Gamma_t^2 p_t)_{ss}
&= 2(\Gamma_t')^2 p_t+2\Gamma_t\Gamma_t''p_t
  +4\Gamma_t\Gamma_t'\partial_s p_t+\Gamma_t^2 p_{ss}.
\end{align*}
Substituting into~\eqref{eq:gamma_group}, almost all terms cancel:
\begin{align*}
&\tfrac{1}{2}\bigl[2(\Gamma_t')^2 p_t+2\Gamma_t\Gamma_t''p_t
  +4\Gamma_t\Gamma_t'\partial_s p_t+\Gamma_t^2 p_{ss}\bigr]
+\tfrac{1}{2}\Gamma_t^2 p_{ss}
-\Gamma_t\bigl[\Gamma_t''p_t+2\Gamma_t'\partial_s p_t+\Gamma_t p_{ss}\bigr]\\
&\quad= (\Gamma_t')^2 p_t = (\Gamma_t')^2 Y_t.
\end{align*}
Thus the dynamics of $Y_t$ simplify to:
\begin{equation}
\D Y_t
= -\Gamma_t'Y_t\,\D Z_t
+ \left[(\Gamma_t')^2 Y_t
        +\tfrac{1}{2}(\Delta_t^2 p_t)_{ss}
        -(\mu_t p_t)_s\right]\D t.
\end{equation}

\medskip
\noindent\textbf{Step 3: Conservation of mass and complete cancellation.}
Define the transformed density $q_t(x)=Y_t(x)J_t(x)$. By the It\^o product
rule:
\[
\D q_t = Y_t\,\D J_t + J_t\,\D Y_t + \D\langle Y,J\rangle_t.
\]
The cross-variation uses only the martingale parts:
$\D\langle Y,J\rangle_t=(-\Gamma_t'Y_t)(\Gamma_t'J_t)\,\D t
=-(\Gamma_t')^2 q_t\,\D t$.
Assembling the full equation:
\begin{align}
\D q_t
&= Y_t(\Gamma_t'J_t\,\D Z_t)
  +J_t\!\left(-\Gamma_t'Y_t\,\D Z_t
             +\left[(\Gamma_t')^2 Y_t
                    +\tfrac{1}{2}(\Delta_t^2 p_t)_{ss}
                    -(\mu_t p_t)_s\right]\D t\right)
  -(\Gamma_t')^2 q_t\,\D t \nonumber\\
&= 0\cdot\D Z_t
  +J_t\!\left[\tfrac{1}{2}(\Delta_t^2 p_t)_{ss}
             -(\mu_t p_t)_s\right]\D t.
\label{eq:dq_clean}
\end{align}
The stochastic integral and the $(\Gamma_t')^2$ drift corrections vanish
completely.

\medskip
\noindent\textbf{Step 4: Coefficient extraction.}
To rewrite the $s$-derivatives in terms of $x$, we use
$\partial_s=J_t(x)^{-1}\partial_x$ and $p_t=q_t/J_t(x)$. The first-order
term is:
\begin{equation}
-J_t(x)\,\bigl[(\mu_t p_t)_s\bigr](\Phi_t(x))
= -\partial_x\!\left(\frac{\mu_t(\Phi_t(x))}{J_t(x)}\,q_t(x)\right).
\end{equation}
For the second-order term,
\[
\frac{1}{2}J_t(x)\,\bigl[(\Delta_t^2 p_t)_{ss}\bigr](\Phi_t(x))
= \frac{1}{2}\partial_x\!\left(
    \frac{1}{J_t(x)}\partial_x\!\left(
      \frac{\Delta_t^2(\Phi_t(x))}{J_t(x)}q_t(x)
    \right)
  \right).
\]
Writing $\mathcal{A}_t(x):=\Delta_t^2(\Phi_t(x))$ for brevity, this becomes
\[
  \frac{1}{2}\partial_x\!\left(
    \frac{1}{J_t(x)}\partial_x\!\left(
      \frac{\mathcal{A}_t(x)}{J_t(x)}q_t(x)
    \right)
  \right).
\]
Since $H_t(x)=\partial_x J_t(x)$, a direct expansion gives
\begin{align*}
  \frac{1}{2}\partial_x\!\left(
    \frac{1}{J_t(x)}\partial_x\!\left(
      \frac{\mathcal{A}_t(x)}{J_t(x)}q_t(x)
    \right)
  \right)
  &=
  \frac{1}{2}\partial_{xx}\!\left(
    \frac{\mathcal{A}_t(x)}{J_t(x)^2}q_t(x)\right)
  +\frac{1}{2}\partial_x\!\left(
    \frac{\mathcal{A}_t(x) H_t(x)}{J_t(x)^3}q_t(x)\right).
\end{align*}
Substituting this together with the first-order term
into~\eqref{eq:dq_clean} yields
\begin{equation}\label{eq:final_pde}
\partial_t q_t(x)
= -\partial_x\!\left(\left[\frac{\mu_t(\Phi_t(x))}{J_t(x)}
                           -\frac{1}{2}\frac{\Delta_t^2(\Phi_t(x)) H_t(x)}
                                            {J_t(x)^3}
                     \right]q_t(x)\right)
+ \tfrac{1}{2}\partial_{xx}\!\left(\frac{\Delta_t^2(\Phi_t(x))}{J_t(x)^2}
  \,q_t(x)\right),
\end{equation}
which defines the pathwise coefficients $\tilde{\mu}_t(x)$ and
$\widetilde{\Delta}_t^2(x)$ stated in
\eqref{eq:Delta_tilde}--\eqref{eq:mu_tilde}. This completes the proof of
Theorem~\ref{thm:main}.\qed

\section{Interpretation, well-posedness, and calibration}
\label{sec:interpretation_calibration}

\begin{remark}[Classical LSV versus rough LSRV]\label{rem:classical_vs_rough}
In the classical LSV diffusion setting, where the volatility factor is
itself an It\^o diffusion and the common-noise driver is Brownian, stronger
robustness results of Clark--Crisan type are available under additional
regularity assumptions ensuring continuity of the filter as a functional of
the observation path. For the motivating It\^o--Wentzell/filtering viewpoint
in the classical LSV setting, see the Lucic~\cite{Lucic1}.
The present article does not revisit that classical theory in full generality.
Its focus is the structural part of the argument that remains valid in the
rough/Volterra setting: the It\^o--Wentzell stochastic-flow transformation
removes the common-noise transport from the conditional-density SPDE and
yields, for each fixed common environment, a pathwise Fokker--Planck type
PDE with random coefficients. In the rough setting, establishing full
Clark--Crisan robustness would require additional analysis in an appropriate
rough-path or Volterra-path topology.
\end{remark}

\begin{remark}[Well-posedness of the pathwise PDE]\label{rem:wellposed}
For each fixed sample path $\omega$, the path $t\mapsto V_t(\omega)$ is a
deterministic, locally bounded, measurable function. The transformed
coefficients $\widetilde\Delta_t^i(x)$ and $\tilde\mu_t^i(x)$ in
equation~\eqref{eq:pathwise_pde} are therefore deterministic functions of
$(t,x)$ after fixing the common-environment realization and the associated
stochastic flow: measurable in~$t$, and with the finite spatial
regularity supplied by Assumption~\ref{ass:coeffs}(i) and~\ref{ass:coeffs}(ii),
together with the corresponding regularity of the stochastic flow and its
spatial derivatives. In general one should not claim more than this local
$C^2/C^1$ type regularity without stronger assumptions.
Thus the transformed equation may be viewed as a random PDE which becomes,
pathwise, a deterministic Fokker--Planck type PDE with path-dependent
coefficients.

In rough Heston, the diffusion coefficient of the transformed
PDE~\eqref{eq:pathwise_pde} is $\widetilde\Delta_t^2(x)=(1-\rho^2)\,x^2
V_t$, which degenerates whenever $V_t=0$. Recent results on the fractional
Volterra square-root process establish that $V_t$ may reach zero with
positive probability before any fixed horizon
$t>0$~\cite{friesen2026boundary,bourdon2026boundary}. Consequently, a
Feller-type condition cannot render zero globally inaccessible in canonical
rough Heston, and globally uniform ellipticity of~\eqref{eq:pathwise_pde}
cannot be assumed. However, for $V_0>0$ and $|\rho|<1$, continuity of $V$
and positivity of $V_0$ imply $\int_0^t V_s\,\D s>0$ for every $t>0$, so
the explicit lognormal representation of Section~\ref{sec:rH_easier} remains
valid.

Equation~\eqref{eq:pathwise_pde} should therefore be interpreted in one of
the following senses, depending on the application:
\begin{enumerate}
\item[\textup{(a)}] \emph{Degenerate parabolic theory}, allowing the
      diffusion coefficient to vanish at the boundary;
\item[\textup{(b)}] \emph{Weak or distributional formulation}, in which
      the equation~\eqref{eq:pathwise_pde} is tested against smooth
      compactly supported functions $\varphi\in C_c^\infty((0,\infty))$,
      and $\partial_t q$ is understood almost everywhere in $t$ or in the
      sense of distributions. A weak or distributional formulation remains
      meaningful under the assumptions of this paper. Existence, uniqueness
      and regularity in the degenerate case require additional
      model-specific coercivity, growth and boundary hypotheses beyond
      those imposed here;
\item[\textup{(c)}] \emph{Local uniform parabolicity}: on any space--time
      cylinder $Q=[t_0,t_1]\times[a,b]$ with $0<a<b<\infty$ and
      $\inf_{(t,x)\in Q}\widetilde\Delta_t^2(x)>0$, the coefficients are
      bounded and the operator is uniformly parabolic. Standard weak
      divergence-form or $L^p$ parabolic theory therefore applies under
      the corresponding coefficient hypotheses~\cite{LSU1968}. Classical
      Schauder regularity requires additional continuity or H\"older
      regularity in time. In pure rough Heston this condition follows on
      $Q=[t_0,t_1]\times[a,b]$, $a>0$, from $|\rho|<1$ and
      $\inf_{t\in[t_0,t_1]}V_t>0$; or
\item[\textup{(d)}] In the \emph{pure rough Heston case}
      ($L\equiv 1$, $f=\sqrt{\,\cdot\,}$), the explicit lognormal
      representation derived in Section~\ref{sec:rH_easier} bypasses the
      need for abstract parabolic theory entirely.
\end{enumerate}
When $|\rho|=1$ the idiosyncratic diffusion vanishes identically,
$\widetilde\Delta_t^2\equiv 0$, and the transformed equation ceases to be
parabolic; the conditional law of $S_t$ given the common environment is then
a Dirac measure rather than a smooth density. Any smooth-density or classical
well-posedness claim should therefore be understood under the additional
hypothesis $|\rho|<1$, which is assumed throughout the density-based
analysis beginning with Section~\ref{sec:LSRV}. The boundary behaviour at
$s=0$ depends on the selected weak or classical solution framework; in the
pure rough Heston case with $S_0>0$, the explicit lognormal formula provides
a direct description on $(0,\infty)$.
\end{remark}

\subsection{Regularity considerations}

When we involve a rough volatility component, we will have a further loss of
time regularity in the final PDE coefficients. The preceding observation
should be understood at a structural level. The It\^o--Wentzell change of
variables itself is unaffected by the rough nature of the volatility process.
However, the coefficients of the transformed pathwise PDE inherit the
temporal regularity of the underlying volatility factor.

For classical stochastic volatility models such as Heston, the resulting
coefficients are already random and sample paths are typically only H\"older
continuous in time with exponent strictly smaller than $1/2$. In rough
volatility models the corresponding H\"older exponent is typically bounded by
the Hurst parameter $H<1/2$, leading to an additional loss of temporal
regularity. Consequently, while the derivation of the pathwise PDE remains
valid under Assumptions~\ref{ass:coeffs}--\ref{ass:density}, the well-posedness and regularity theory
for the transformed equation is discussed in Remark~\ref{rem:wellposed}
above: $L^\infty$ measurability in time suffices for the pathwise algebraic
reduction itself; well-posedness of the resulting PDE requires the additional
boundedness, coercivity, growth and boundary conditions appropriate to the
selected solution space.

\subsection{Numerical implications and Rao--Blackwellized calibration}
\label{sec:numerical}

For the calibration discussion we additionally assume that
\[
  \exp\!\left(-\int_0^t(r_u-d_u)\,\D u\right)S_t
\]
is a true martingale on $[0,T]$, and that the target call-price surface
has the regularity and boundary behaviour required for the standard Dupire
formula.

While the analytic transformation survives intact, the numerical
implementation must be adapted to account for the correlation $\rho$ between
the spot noise $B_t$ and the rough volatility noise $W_t=Z_t$.

In the standard model, conditioning on a path of $Z_t^i$ might allow for
certain decoupled simulation techniques. In the rough setting, the volatility
factor is Volterra-driven and depends nonlocally on the full past of its
common Brownian driver. Because $B_t$ and $W_t=Z_t$ are correlated,
simulation of the common noise path $Z_t^i$ must be carried out consistently
with the corresponding path of $V_t^i$.

Therefore, the stochastic integral is encoded into the random flow, but the
generation of the underlying randomness requires a joint scheme. The pathwise
filtering sequence is modified as follows:
\begin{enumerate}

\item \textbf{Joint Rough Simulation:} Simulate the correlated joint paths
      $(Z^{i},V^{i})$. For Gaussian Volterra models such as rough Bergomi,
      the hybrid scheme of Bennedsen, Lunde and
      Pakkanen~\cite{bennedsen2017hybrid} applies directly. For the
      nonlinear rough Heston variance equation, dedicated discretizations
      are required: Euler-type Volterra schemes, low-dimensional Markovian
      approximations, integrated-variance schemes, or other discretizations
      specifically designed for Volterra square-root equations. Cholesky
      decomposition of the joint covariance matrix is appropriate for
      sampling correlated Gaussian vectors but is not by itself a
      discretization of the nonlinear rough Heston variance equation.

\item \textbf{Flow Evolution:} Evolve the random flow components
      $(\Phi^{i},J^{i},H^{i})$ along the path $Z^i$, using $V^i$ as a
      deterministic forcing term.

\item \textbf{Pathwise PDE Update:} After fixing the common-environment
      realization, solve the resulting deterministic Fokker--Planck type
      PDE~\eqref{eq:pathwise_pde} for $q_t^i(x)$ using standard finite
      difference methods or Assumed Density/Galerkin projections. The
      temporal grid must be sufficiently fine to capture the rough
      continuous variations of the coefficients.
      \begin{remark}
      The reduced temporal regularity of the coefficients, typically
      governed by the Hurst parameter $H<\tfrac{1}{2}$, can be expected to
      lower the achievable temporal order of convergence of standard
      time-stepping schemes. In practice this necessitates a sufficiently
      fine temporal grid, and implicit or semi-implicit discretizations are
      often preferable for stability.
      \end{remark}

\item \textbf{Coordinate Inversion:} Recover $p_t^i(s)$ from $q_t^i(x)$
      using the inverse flow. In particular, for a strike $K$, define the
      preimage
      \[
        x_K^i(t):=(\Phi_t^i)^{-1}(K).
      \]
      Since $q_t^i(x)=p_t^i(\Phi_t^i(x))\,J_t^i(x)$, we obtain
      \begin{equation}\label{eq:pt_from_q}
        p_t^i(K)
        = \frac{q_t^i(x_K^i(t))}{J_t^i(x_K^i(t))}.
      \end{equation}

\item \textbf{Rao--Blackwellized Calibration Ratio:} Intuitively, by the
      law of total expectation, the Dupire calibration constraint involves
      the conditional moment $\EE[f^2(V_t)\mid S_t=K]$.

Set
\[
  \bar p_t(K):=\mathbb E[p_t(K)],
  \qquad
  n_t(K):=\mathbb E[f^2(V_t)p_t(K)].
\]
For Lebesgue-almost every $K$ such that $\bar p_t(K)>0$,
\[
  \Theta(t,K):=\frac{n_t(K)}{\bar p_t(K)}
\]
is a version of $\mathbb E[f^2(V_t)\mid S_t=K]$.
If $n_t$ and $\bar p_t$ admit continuous versions, the same ratio
defines the corresponding pointwise version wherever $\bar p_t(K)>0$.

Approximating the outer expectations by independent common-environment
realisations $(Z^i,V^i)$ and evaluating the corresponding conditional
densities yields the following Rao--Blackwellized estimator in transformed
coordinates:
      \begin{equation}\label{eq:RB_theta_transformed}
        \hat{\Theta}(t,K)
        =
        \frac{\displaystyle \sum_{i=1}^{N}
          f^{2}(V_t^i)\,
          \dfrac{q_t^i(x_K^i(t))}{J_t^i(x_K^i(t))}}
        {\displaystyle \sum_{i=1}^{N}
          \dfrac{q_t^i(x_K^i(t))}{J_t^i(x_K^i(t))}},
      \end{equation}
      assuming $|\rho|<1$. Equivalently, using~\eqref{eq:pt_from_q},
      \begin{equation}\label{eq:RB_theta_density}
        \hat{\Theta}(t,K)
        =
        \frac{\displaystyle \sum_{i=1}^{N}
          f^{2}(V_t^i)\,p_t^i(K)}
        {\displaystyle \sum_{i=1}^{N} p_t^i(K)},
      \end{equation}
      assuming $|\rho|<1$. Formula~\eqref{eq:RB_theta_density} is the
      more natural expression for interpretation, while
      \eqref{eq:RB_theta_transformed} is computationally preferable since
      it evaluates the ratio directly in transformed coordinates. In an
      actual calibration loop, however, the quantities $p_t^i$, $q_t^i$,
      and $\Phi_t^i$ all depend on the current leverage function $L$, so
      the local-leverage update is to be understood as a fixed-point/Picard
      step rather than a one-shot closed formula:
      \begin{equation}\label{eq:leverage_update}
        \hat{L}^{2}(t,K)
        = \frac{\sigma_{\mathrm{Dup}}^{2}(t,K)}
               {\hat{\Theta}(t,K)}.
      \end{equation}
      Thus the rough extension leaves the calibration formula structurally
      unchanged: the only modification is that the quantities entering
      $\hat{\Theta}(t,K)$ are now obtained from the transformed pathwise
      PDE with rough, path-dependent coefficients, and from the joint
      simulation of $(Z^i,V^i)$. The finite-$N$ ratio estimator is
      self-normalized and converges consistently as $N\to\infty$ under iid
      sampling of the common-environment realisations and appropriate
      integrability, provided $\EE[p_t(K)]>0$; it should be viewed as a
      natural consistent approximation rather than as an exactly unbiased
      estimator.
\end{enumerate}

The formulas~\eqref{eq:RB_theta_transformed}--\eqref{eq:leverage_update}
show that the Rao--Blackwellized calibration mechanism survives unchanged at
the structural level under the rough extension. The It\^o--Wentzell
transformation removes the common-noise transport exactly as in the classical
LSV case, while the rough/non-Markovian nature of the volatility factor
enters only through the random coefficients of the transformed PDE and the
joint simulation step.

\section{The rough Heston specialization}\label{sec:rH_easier}

Rough Heston is not merely covered by the general framework; in some
respects it is structurally simpler than a generic local stochastic
volatility model because the transport coefficient is affine in the spatial
variable.

We restore the deterministic rates $r$ and dividend yields $d$ appearing in
the general LSRV specification, dropping the simplifying zero-drift
convention used in Section~\ref{sec:rHestonintro}.

For rough Heston one has $L(t,S)\equiv1$ and $f(V)=\sqrt{V}$, so that
$\Gamma_t(s)=\rho s\sqrt{V_t}$. The transport coefficient is affine in the
spatial variable $s$, and the associated stochastic flow satisfies
\[
\D\Phi_t(x) = \rho\,\Phi_t(x)\sqrt{V_t}\,\D Z_t.
\]
This geometric SDE has the explicit solution (obtained by applying It\^o's
formula to $\log \Phi_t$),
\begin{equation}\label{eq:rH_flow_explicit}
\Phi_t(x)
= x\exp\!\left(
    \rho\int_0^t\!\sqrt{V_s}\,\D Z_s
    -\tfrac{\rho^2}{2}\int_0^t V_s\,\D s
  \right),
\end{equation}
where $M_t := \exp\!\bigl(\rho\int_0^t\!\sqrt{V_s}\,\D Z_s -
\tfrac{\rho^2}{2}\int_0^t V_s\,\D s\bigr)$, so that
\begin{equation}\label{eq:rH_JH}
J_t(x)=M_t,\qquad x\ge 0,
\qquad
H_t(x)=0.
\end{equation}
The identity at $x=0$ follows directly by differentiating the linear map
$x\mapsto xM_t$.

Substituting~\eqref{eq:rH_JH} into~\eqref{eq:Delta_tilde}--\eqref{eq:mu_tilde}
gives the simplified coefficients
\begin{equation}\label{eq:rH_coeffs}
\widetilde{\Delta}_t^2(x) = (1-\rho^2)\,x^2\,V_t,
\qquad
\tilde{\mu}_t(x) = (r_t-d_t)\,x,
\end{equation}
which is a strong sanity check: the transformed PDE carries the residual
idiosyncratic variance $(1-\rho^2)V_t$ and the standard risk-neutral drift,
exactly as expected after removing the common-noise transport.

Hence the spatial regularity assumptions required for the flow construction
are immediate in this affine case. The only additional complexity relative to
the classical Heston case is the reduced temporal regularity of the random
coefficient $V_t$.

If we are prepared to accept the presence of a local component $L(t,S)$,
even in that case the result generalizes, as we have seen, although we do not
benefit from the rHeston simplification above.

In fact, in the pure rough Heston case the transformed
PDE~\eqref{eq:pathwise_pde} admits an explicit closed-form solution,
providing a stronger verification than the coefficient sanity
check~\eqref{eq:rH_coeffs} above. Define the common stochastic integral
$I_t$, the total integrated variance $\mathcal{V}_t$, the accumulated
idiosyncratic variance $A_t$, and the integrated risk-neutral drift $R_t$ by
\begin{equation}\label{eq:rH_IVt}
I_t := \int_0^t \sqrt{V_u}\,\D Z_u,
\qquad
\mathcal{V}_t := \int_0^t V_u\,\D u,
\qquad
A_t := (1-\rho^2)\,\mathcal{V}_t,
\qquad
R_t := \int_0^t (r_u - d_u)\,\D u,
\end{equation}
so that $M_t = \exp\!\bigl(\rho I_t -
\tfrac{\rho^2}{2}\mathcal{V}_t\bigr)$. Conditional on the
common-environment realization $(Z^i,V^i)$, both $A_t$ and $R_t$ are
deterministic functions of time for each scenario~$i$. Since $\Phi_t(x) = x
M_t$ with $J_t(x) = M_t$ and $H_t(x)=0$, the transformed process $X_t :=
S_t/M_t$ satisfies, by It\^o's formula,
\begin{equation}\label{eq:rH_Xt_sde}
\frac{\D X_t}{X_t}
= (r_t - d_t)\,\D t
+ \sqrt{1-\rho^2}\,\sqrt{V_t}\,\D\zeta_t.
\end{equation}
Since $\zeta$ is independent of the common filtration
$\widehat{\mathcal C}_\infty$ (Remark~\ref{rem:filtration}), and $V_t$ is
$\widehat{\mathcal C}_t$-measurable, the conditional law of $\log X_t$ given
$\widehat{\mathcal C}_t$ is Gaussian with mean $\log S_0 + R_t -
\tfrac{1}{2}A_t$ and variance $A_t$. Whenever $A_t > 0$, which holds almost
surely for every $t>0$ when $V_0>0$ by continuity of $V$ (see
Remark~\ref{rem:wellposed}), the transformed density is the lognormal
\begin{equation}\label{eq:rH_qt_explicit}
q_t(x)
= \frac{1}{x\sqrt{2\pi A_t}}
  \exp\!\left(
    -\frac{
      \bigl(\log(x/S_0) - R_t + \tfrac{1}{2}A_t\bigr)^2
    }{2A_t}
  \right),
\qquad x > 0,
\end{equation}
and the original conditional density is recovered via the inverse flow
$s = \Phi_t(x) = xM_t$:
\begin{equation}\label{eq:rH_pt_explicit}
p_t(s)
= \frac{1}{M_t}\,q_t\!\left(\frac{s}{M_t}\right)
= \frac{1}{s\sqrt{2\pi A_t}}
  \exp\!\left(
    -\frac{
      \Bigl(\log(s/S_0) - R_t - \rho I_t
            + \tfrac{1}{2}\mathcal{V}_t\Bigr)^2
    }{2A_t}
  \right),
\qquad s>0.
\end{equation}

We now verify Assumption~\ref{ass:density} rigorously in pure rough
Heston. Fix $\tau>0$ and a compact set $K\Subset(0,\infty)$. Since $V_0>0$,
$V$ is continuous and nonnegative, and $|\rho|<1$, we have
\[
  A_t=(1-\rho^2)\int_0^t V_u\,\D u \ge A_\tau>0,
  \qquad t\in[\tau,T],
\]
almost surely. Write
\[
  p_t(s)=G\bigl(s,I_t,\mathcal V_t,R_t\bigr),
\]
where $G$ is the smooth function on
$(0,\infty)\times\mathbb R\times(0,\infty)\times\mathbb R$ defined by the
right-hand side of~\eqref{eq:rH_pt_explicit}. Since
\[
  \D I_t=\sqrt{V_t}\,\D Z_t,
  \qquad
  \D\mathcal V_t=V_t\,\D t,
  \qquad
  \D R_t=(r_t-d_t)\,\D t,
\]
It\^o's formula yields, for each fixed $s>0$,
\[
  \D p_t(s)=a_t(s)\,\D t+g_t(s)\,\D Z_t,
\]
with
\[
  g_t(s)=\sqrt{V_t}\,\partial_i G\bigl(s,I_t,\mathcal V_t,R_t\bigr),
\]
and
\[
  a_t(s)
  =
  V_t\,\partial_v G\bigl(s,I_t,\mathcal V_t,R_t\bigr)
  +(r_t-d_t)\,\partial_r G\bigl(s,I_t,\mathcal V_t,R_t\bigr)
  +\tfrac12 V_t\,\partial_{ii}G\bigl(s,I_t,\mathcal V_t,R_t\bigr).
\]

For completeness, write
\[
  B_t(s):=\log(s/S_0)-R_t-\rho I_t+\tfrac12\mathcal V_t,
  \qquad
  A_t=(1-\rho^2)\mathcal V_t,
\]
so that
\[
  p_t(s)=\frac{1}{s\sqrt{2\pi A_t}}
  \exp\!\left(-\frac{B_t(s)^2}{2A_t}\right).
\]
Then
\[
  \partial_i G\bigl(s,I_t,\mathcal V_t,R_t\bigr)
  =
  \frac{\rho\,B_t(s)}{A_t}\,p_t(s),
\]
because $\partial_i B_t(s)=-\rho$. Also,
\[
  \partial_s B_t(s)=\frac{1}{s},
\]
and therefore
\[
  \partial_s p_t(s)
  =
  -\frac{1}{s}p_t(s)-\frac{B_t(s)}{A_t\,s}p_t(s).
\]
Hence
\begin{align*}
-\partial_s\bigl(\rho s\sqrt{V_t}\,p_t(s)\bigr)
&=
-\rho\sqrt{V_t}\,\bigl(p_t(s)+s\,\partial_s p_t(s)\bigr)\\
&=
-\rho\sqrt{V_t}\left(
p_t(s)+s\left[-\frac{1}{s}p_t(s)-\frac{B_t(s)}{A_t\,s}p_t(s)\right]
\right)\\
&=
\rho\sqrt{V_t}\,\frac{B_t(s)}{A_t}\,p_t(s)
=
\sqrt{V_t}\,\partial_i G\bigl(s,I_t,\mathcal V_t,R_t\bigr).
\end{align*}
Thus
\[
  g_t(s)= -\partial_s\bigl(\rho s\sqrt{V_t}\,p_t(s)\bigr).
\]

Since the adapted processes $I$, $\mathcal V$, $R$, and $V$ are continuous,
the coefficient fields $a_t(s)$ and $g_t(s)$ are progressively measurable in
$t$. Moreover, for each $\omega$, the paths
$t\mapsto I_t(\omega)$, $t\mapsto\mathcal V_t(\omega)$, and
$t\mapsto R_t(\omega)$ have compact ranges on $[\tau,T]$. Hence, for almost
every $\omega$, the set
\[
  \mathcal K_\omega
  :=
  K\times I_{[\tau,T]}(\omega)\times \mathcal V_{[\tau,T]}(\omega)\times
  R_{[\tau,T]}(\omega)
\]
is compact and contained in
\[
  (0,\infty)\times\mathbb R\times[A_\tau(\omega)/(1-\rho^2),\infty)\times
  \mathbb R.
\]
Because $G$ is smooth on this domain and the third coordinate stays bounded
away from zero, the derivatives of $G$ entering the formulas for $p$, $a$,
and $g$ are bounded on $\mathcal K_\omega$. It follows that
\[
  \sup_{t\in[\tau,T]}\|p_t\|_{C^2(K)}<\infty
  \qquad\text{a.s.},
\]
and
\[
  \int_\tau^T
  \Bigl(
    \|a_u\|_{C^0(K)}+\|g_u\|_{C^1(K)}^2
  \Bigr)\,\D u<\infty
  \qquad\text{a.s.}
\]
Comparing this semimartingale decomposition with the weak
conditional-density SPDE and using uniqueness of the continuous
semimartingale decomposition gives
\[
  a_t(s)
  =
  \tfrac12\partial_{ss}\bigl(\Sigma_t^2(s)p_t(s)\bigr)
  -\partial_s\bigl(\mu_t(s)p_t(s)\bigr).
\]
Since the coefficients and density are spatially smooth on
$[\tau,T]\times K$, this equality holds there in the local classical
sense. Hence Assumption~\ref{ass:density} holds in pure rough Heston.

The conditioned geometric SDE~\eqref{eq:rH_Xt_sde} has a unique strong
solution, and formula~\eqref{eq:rH_qt_explicit} is the density of its law.
It therefore provides the canonical probability-measure solution of the
transformed Fokker--Planck equation. To make the uniqueness argument
transparent, set $c_t := (1-\rho^2)V_t$ and $b_t := r_t - d_t$, and define
the log-density
\[
  \bar{q}_t(y) := e^y\,q_t(e^y),
\]
which is the density of $\log X_t$ with respect to $\D y$. Then
\begin{equation}\label{eq:log_heat}
  \partial_t\bar{q}_t(y)
  = -\partial_y\!\left[\left(b_t - \tfrac{1}{2}c_t\right)\bar{q}_t(y)\right]
  + \tfrac{1}{2}c_t\,\partial_{yy}\bar{q}_t(y).
\end{equation}
Let
\[
  m_t:=b_t-\tfrac12c_t,
  \qquad
  \beta_t:=\int_0^t m_u\,\D u,
\]
and define
\[
  u_t(z):=\bar q_t(z+\beta_t).
\]
Then
\[
  \partial_tu_t(z)
  =
  \tfrac12c_t\,\partial_{zz}u_t(z).
\]
Taking Fourier transforms in $z$ gives
\[
  \widehat u_t(\xi)
  =
  \exp\!\left(-\tfrac12\xi^2A_t\right)
  \widehat u_0(\xi),
  \qquad
  A_t=\int_0^t c_u\,\D u.
\]
Hence the solution is uniquely determined by the Gaussian transition kernel,
even if $A$ has flat portions. This gives uniqueness in the class of
narrowly continuous probability solutions of the transformed Fokker--Planck
equation~\eqref{eq:pathwise_pde} in pure rough Heston, as long as we
consider the probability solutions on $(0,\infty)$ whose logarithmic
pushforwards are narrowly continuous probability-measure solutions on
$\mathbb{R}$.

It provides a closed-form benchmark for the numerical implementation of
Step~3 in Section~\ref{sec:numerical}, and constitutes a stronger sanity
check than~\eqref{eq:rH_coeffs}: not only are the coefficients correct, but
the full solution is consistent with the direct probabilistic argument.

\section{Conclusion}\label{sec:conclusion}

The paper shows that the pathwise filtering transformation of the
conditional-density equation survives under local stochastic rough
volatility, provided the common filtration is chosen so that $Z$ remains a
Brownian motion and the required measurability, predictability and
spatial-regularity assumptions hold. The common-noise transport can then be
removed by an It\^o--Wentzell stochastic flow, leaving for each fixed
common-environment realization a deterministic Fokker--Planck type PDE
without an explicit rough driver, with path-dependent coefficients.

The non-Markovian memory is localized to the generation of the common
environment and to the resulting time-dependent coefficients; it does not
introduce a new stochastic transport term after transformation. This makes
the transformed equation compatible with standard pathwise PDE methods and,
in the calibration setting, yields the Rao--Blackwellized leverage update in
a form directly analogous to the classical LSV case.

For pure rough Heston, the transport coefficient is affine, the stochastic
flow is explicit, the transformed coefficients simplify, and the conditional
density can be written in closed form. This solvable case provides a useful
verification of the general transformation and a benchmark for numerical
implementation.

A full Clark--Crisan robustness theorem in rough settings is not proved here.
Establishing continuity of the filter with respect to the observation path in
a genuinely rough-path or Volterra-path topology remains an interesting
direction for future work.

\end{document}